\documentclass{article}
\usepackage{amsmath}
\usepackage{amstext}
\DeclareMathOperator{\ad}{ad} \DeclareMathOperator{\Imag}{Imag}
\DeclareMathOperator{\Index}{Index}
 
\DeclareMathOperator{\ch}{ch} \DeclareMathOperator{\td}{td}
\DeclareMathOperator{\e}{e} \DeclareMathOperator{\rank}{rank}
 \font\cero=cmss10 scaled 1728
\font\uno=cmssbx10 scaled 1200 
\usepackage{graphicx}
\begin{document}
\begin{flushleft}
{\cero  Dimension of the moduli space and Hamiltonian analysis of BF field theories} \\
\end{flushleft}
{\sf R. Cartas-Fuentevilla},
{\sf A. Escalante-Hernandez} \\
 {\it Instituto de F\'{\i}sica,
Universidad Aut\'onoma de Puebla,
Apartado postal J-48 72570 Puebla Pue., M\'exico}; \\
\noindent{\sf J. Berra-Montiel} \\
 {\it Facultad de Ciencias F\'{\i}sico Matem\'{a}ticas, Universidad Au\-t\'o\-no\-ma de Puebla,
 Apartado postal 1152, 72001 Puebla, Pue., M\'exico.}
 \\ \\


\noindent{\uno Abstract} \vspace{.5cm}\\
By using the Atiyah-Singer theorem through some
similarities with the instanton and the anti-instanton moduli
spaces,
the dimension of the moduli space for two and four-dimensional BF
theories valued in different background manifolds and gauge groups
scenarios is determined. Additionally, we develop Dirac's canonical analysis for a
four-dimensional modified BF theory, which reproduces the
topological YM theory. This framework will allow us to understand
the local symmetries, the constraints, the extended Hamiltonian and
the extended action of the theory.
\\

PACS numbers: . \\ \\
 \noindent {\uno I. Introduction and motivations}\vspace{.5em}\\
The BF formalism \cite{1,2,3} in arbitrary space or space-time
dimensions has shown profound relationships between topological
field theory and quantum field theory, from a simplified version of
general relativity \cite{4,5}, to a new formulation of Yang-Mills (YM)
theory \cite{6,7,8,9,10,11}. In the so called first-order formulation, YM
theory can be viewed as a perturbative expansion in the coupling
constant $g$ around the pure topological BF theory; additionally the
BF first-order formulation is {\it on shell} equivalent to the usual
(second-order) YM theory. In this context, both formulations of the
theory possess the same perturbative quantum properties \cite{12,13}.
Furthermore, the Feynman rules, the structure of one loop divergent
diagrams, and renormalization have been studied, and the equivalence
of the {\it uv}-behavior of both formulations has been verified
\cite{14}. However, in spite of these developments, there exist
certain basic aspects poorly understood in the specific case of
four-dimensional BF theories, which is ironic as already mentioned
by J. Baez \cite{4,5}, since four-dimensional gauge theories are the
main motivation, and the obvious subject of research; this is in
part due to certain technical complications that the
four-dimensional case has in relation to low-dimensional scenarios.
It is surprising that, for example, the dimension of the
corresponding moduli spaces of four-dimensional BF theory have not
been determined for a general base manifold. Especifically, the
natural question to be asked is whether there exists any
relationship between four-dimensional YM instantons moduli space and
the corresponding moduli space of four-dimensional BF theory. In
fact, such a relationship there exists in the case of BF fields on a
Riemann surface, and the two-dimensional YM instantons on it 
(\cite{1},\cite{15}, and references therein). In
the present work we attempt to explore the BF moduli space using
some similarities between the BF complex and YM instanton complex in
four dimensions, employing as the main tool the Atiyah-Singer
theorem.

\noindent On the other hand, the instanton moduli space can be
considered as a starting point for quantizing a field theory around
a non-discrete space of classical minima, where the functional
integration over the moduli space is treated non-perturbatively,
whereas the integration over the quantum fluctuations
``perpendicular" to the moduli space can be treated perturbatively
\cite{16,17}. This scheme has been called the ``topological embedding",
where the essential idea is that the moduli space around which the
field theory is studied perturbatively possesses an enhanced gauge
symmetry, the topological invariance; it is here where the
topological BF theory fits naturally within the {\it topological
embedding} setting since it constitutes the topological sector
around which QCD or general relativity can be expanded. Therefore,
it is crucial to obtain geometrical, topological and physical
information about the moduli space of the BF theory, particularly in
the case of four dimensions.

\noindent In order to obtain relevant information on the moduli
space of the BF theory, it is possible to draw on the closely
relation between four-dimensional BF moduli space and
four-dimensional YM instantons and anti-instantons moduli spaces;
this relation is based on the flatness condition, which is one of
the equations of motion of the BF theory. It follows that
connections meeting self-duality and anti-self-duality conditions
simultaneously, namely, the connections belonging to the
intersection of the instanton and the anti-instanton moduli spaces,
are hence solutions of the flatness condition on the curvature. The
space formed by these solutions up to gauge transformations is known
as the moduli space of BF; as in the case of YM instantons
corresponds generally to a finite-dimensional smooth manifold. This
manifold is usually non-compact, partly due to conformal invariance
of the equations of motion, leading to technical difficulties in the
applications. It turns out that in order to define a well behaved
gauge field theory, one needs to regularize the model to avoid
problems with reducible connections. Reducible connections are
source of great difficulty in making sense to the quantization of
gauge field theories in general; the problem is that at reducible
connections the path integrals related to partition functions
diverge. This regularization amounts to considering a modified
four-dimensional BF theory, reproducing in the limit the usual BF
theory and the four-dimensional topological YM theory
\cite{18}. The main reason to use a modified version of the BF
theory as we shall see within the Hamiltonian analysis, is that
unlike the BF-YM theory, it shares the same gauge symmetries with
the usual BF theory, referring with particular emphasis on
diffeomorphisms. However, both cases can be treated perturbatively
around the moduli space defined for flat connections within the
\textit{topological embedding} mentioned above.

\noindent On the other hand, all the information we need to
calculate the dimension of the moduli space using the Atiyah-Singer
theorem, is given by the equations of motion and the local
symmetries of the theory. In order to know these local properties
the Hamiltonian analysis is performed, identifying the relevant
symmetries of the theory such as the extended action, the extended
Hamiltonian and the gauge transformations, with particular emphasis
to the latest since they allow us to build the elliptic complex
which will provide all information about the global degrees of
freedom of the theory under study.

\noindent In the next sections we outline the basic aspects of a BF
theory and the index theorem calculus given by the historic works by
Atiyah {\it et al} \cite{19}, but following the detailed calculations
given in \cite{20}; this will allow us to extend certain
aspects and to modify other ones, for adapting to the special
features of the BF moduli space. In Section IV, as a simple example
the dimension of the moduli spaces for (non-Abelian) two-dimensional
BF theory on a Riemann surface are determined in terms of the
topological and geometrical invariants of the base manifold and the
gauge bundle. In Section V following the example of the previous
section, we characterize the dimension of moduli spaces for a
four-dimensional BF theory and the general expression founded is
used in particular base manifolds and gauge groups scenarios. In
Section VI we present the Hamiltonian analysis for another
BF-topological YM theory. As important results we shall find the
extended action, the extended Hamiltonian and the gauge symmetries
for the theory. In particular we prove that the theory under study
is invariant under diffeomorphisms.
We finish in Section VII with some concluding remarks and prospects. \\

\noindent {\uno II. BF theory in four dimensions}
\vspace{.5em}\\
Let $M$ be an oriented smooth four-dimensional manifold, $B$ a
differential two-form, and $F_{A}$ the curvature induced by a
connection $A$ on a principal bundle over $M$ with structure group
$G$; the BF action is given by \cite{1,2,3}
\begin{equation}
     S_{BF} = \int_{M} Tr B \wedge F_{A};
\end{equation}
this non-Abelian action has the symmetry
\begin{equation}
      A \rightarrow A + d_{A} \Lambda \qquad \textrm{(and then } F_{A} \rightarrow F_{A} + [F_{A}, \Lambda ]), \qquad
      B \rightarrow B + [B, \Lambda ],
\end{equation}
and additionally,
\begin{equation}
     A \rightarrow A, \qquad B \rightarrow B + d_{A} \chi,
\end{equation}
where $\Lambda$ corresponds to an arbitrary 0-form on $M$, and $\chi$ to an arbitrary 1-form; the symmetry (3) requires the Bianchi
identities $d_{A} F = 0$, and is satisfied modulo a total derivative.

The equations of motion obtained from (1) read,
\begin{equation}\label{eq:BF}
     F_{A} = 0, \qquad d_{A} B = 0.
\end{equation}
The action (1) can be obtained in the limit of vanishing coupling $(g \rightarrow 0)$ of the first order formulation of YM
theory given by the action \cite{6,7,8,9}
\begin{equation}\label{5}
     S_{\rm BF-YM} = \int_{M} Tr (i B \wedge F + \frac{g^{2}}{4} B \wedge \ast B),
\end{equation}
where $\ast$ stands for the Hodge-duality operation, and with the gauge symmetry given in (2); it is only in the limit $g\rightarrow 0$
that the second gauge symmetry (3) is present. Furthermore, the equations of motion of the action (5) read
\begin{equation}
     F = i\frac{g^{2}}{2} \ast B, \qquad d_{A} B = 0;
\end{equation}
thus, the substitution of equations (6) into the action (5) leads to the standard $YM$ action
\begin{equation}
     S_{\rm YM} = \frac{1}{g^{2}} \int Tr F \wedge \ast F;
\end{equation}
therefore the BF-YM theory is {\it on-shell} equivalent to YM
theory. Similarly we can find an (on-shell) equivalence between
topological BF-YM theory and topological YM theory considering Eqs.\
(5) (with the symmetry (2)), (6), and (7) with the Hodge-duality
operations removed; thus, the dependence on a metric structure of
$M$ is removed, (see section V). In this case the action (1) is also
the vanishing coupling limit of the topological BF-YM theory. The
action functional (5) and its topological version allow us to
understand YM theory and topological YM theory as {\it perturbative
expansions} in the coupling $g$ around the topological pure BF
theory (1) \cite{10,11}, which defines an authentic topological quantum
field theory\cite{1,2,3}.

Therefore, our fundamental topological sector is given by the pure BF action (1), whose moduli spaces are defined as the spaces of solutions of the corresponding
equations of motion (4) modulo the gauge symmetries (2) and (4). More specifically we can define the $A$-moduli space as the space of (flat) connections satisfying the first of equations (4) modulo the gauge symmetry (2); additionally we define the $B$-moduli space as the space of two-forms satisfying the second of equations (4) modulo the gauge symmetry (3).\\

\noindent {\uno III. Atiyah-Singer index theorem} \vspace{.5em}

Let $M$ be a $n$-dimensional compact smooth manifold without boundary, $\Gamma (E_{p})$ sections of the (complex) vector bundles $E_{p}$ on $M$,
$D_{p}$ differential operators mapping between sections as indicated in the following finite sequence
\begin{eqnarray}\label{complex sequence}
     \!\! & & \!\! \cdots \longrightarrow \Gamma (E_{p-1}) \stackrel{D_{p-1}}{\longrightarrow} \Gamma (E_{p})  \stackrel{D_{p}}{\longrightarrow}
     \Gamma (E_{p+1}) \longrightarrow \cdots \\
     \!\! & & \!\! \cdots \longleftarrow \Gamma (E_{p-1}) \stackrel{D^{\dag}_{p-1}}{\longleftarrow} \Gamma (E_{p})  \stackrel{D^{\dag}_{p}}{\longleftarrow} \Gamma (E_{p+1}) \longleftarrow \cdots \nonumber
\end{eqnarray}
where $D^{+}_{p}$ corresponds to the dual operator of $D_{p}$; if
the Laplacian of the sequence $\Delta_{p} = D^{\dag}_{p} D_{p} +
D_{p-1} D^{\dag}_{p-1}$  is an {\it elliptic} differential operator
\cite{19,21}, then the sequence (8) defines an {\it elliptic
complex} with an {\it index} expressed as
\begin{equation}\label{index}
      \textrm{Index}(E,D) = (-1)^{\frac{n}{2} (n+1)} \int_{M} \frac{\sum_{p} (-1)^{p} ch(E_{p})}{e(T(M))} td (T(M) \otimes C),
\end{equation}
where $ch(E)$ correspond to the Chern characters of the vector
bundles $E$, $td(T(M) \otimes C)$ to the Todd class of the
complexified tangent bundle $T(M) \times C$ of the manifold $M$, and
$e(T(M))$ is the Euler class of the tangent bundle $T(M)$.

It is important to mention that in general the procedure to calculate a moduli space dimension through the Atiyah-Singer index theorem is actually a way to estimate such a dimension, since the dimension may to have unexpected and inadmissible values; a reason is the presence of a nontrivial second cohomology group in the corresponding elliptic complex. In this sense the index calculated corresponds to a {\it virtual} dimension, which will require additional considerations in order to obtain a real dimension. For example it is common the appearance of negative values of the virtual dimension, which will be associated with a empty moduli space; this will be a basic criterion in the present work, as usual in the instanton calculus scenario. \\

\noindent {\uno IV. BF moduli space on a Riemann surface}
\vspace{.5em}

In their own right, gauge theories in two dimensions have for a long
time served as useful laboratories for testing ideas and gaining
insight into the properties of field theories in general,
specifically quantum gauge theories on arbitrary Riemann surfaces.
Our basic concern in this section is with the space or flat
connections (gauge fields) on a compact Riemann surface of genus
\textit{g}, $M=\Sigma_{g}$, and a compact gauge group $G$. A
connection $A$ on a $G$ bundle over $M$, or a gauge field on $M$, is
said to be flat when its curvature tensor $F_{A}$ vanishes,
\begin{equation}\label{eq:3.10}
F_{A}=d\Lambda+\frac{1}{2}[\Lambda,\Lambda]=0.
\end{equation}
Flatness is preserved under gauge transformations $A \rightarrow
A^{U}$ where
\begin{equation}
A^{U}=U^{-1}AU+U^{-1}dU,
\end{equation}
as $F_{A}$ transforms to $U^{-1}F_{A}U$. The moduli space of flat
connections $\mathcal{M}$$_{F}(M,G)$ is the space of gauge
inequivalent solutions to (\ref{eq:3.10}). This means that solutions
to (\ref{eq:3.10}) which are not related by a gauge transformation
are taken to be different points of $\mathcal{M}$$_{F}(M,G)$. On the
other hand, if the solutions are related by a gauge transformation
they are taken to be the same point in $\mathcal{M}$$_{F}(M,G)$,
that is $\{A^{U}\}=\{A\}$.

The usual description of the moduli space in terms of representation
of the fundamental group $\pi_{1}(M)$ of the manifold $M$ \cite{21},
\begin{equation}
\mathcal{M}_{F}(M,G)=Hom(\pi_{1},G)/G,
\end{equation}
that is of equivalent classes of homomorphisms
\begin{equation}
\varphi:\pi_{1}(M)\rightarrow G,
\end{equation}
up to homotopic conjugation. $\pi_{1}(M)$ is made up of loops on the
manifold $M$ with two loops are identified if they can be smoothly
deformed into each other; for example all contractible loops can be
identified. Using homotopy \cite{22} there is a standard
presentation of $\pi_{1}$ in terms of $2g$ generators which are not
independent, since they satisfy relations on the Riemann surface. To
give an index approach we use the Atiyah-Singer theorem allowing us
naturally to make an extension, particularly in the case of four
dimensions where the calculations through homotopy classes could be
rather involve. We now concentrate on the dimension of
$\mathcal{M}_{F}(\Sigma_{g},G)$, which in this case it is known that
is smooth except at singular points which arise at reducible
connections \cite{19}. To achieve this, the natural elliptic complex
to use for our index calculation is
\begin{equation}
0\stackrel{i}{\longrightarrow}\Omega^{0}(\Sigma_{g},\ad
P)\stackrel{d_{A}}{\longrightarrow}\Omega^{1}(\Sigma_{g},\ad
P)\stackrel{d_{A}}{\longrightarrow}\Omega^{2}(\Sigma_{g},\ad
P)\stackrel{d_{A}}{\longrightarrow}0,
\end{equation}
being $\ad P$ the bundle where the gauge group $G$ is defined,
ensuring the correct behavior of flat connections under gauge
transformations. This complex is the corresponding finite sequence
of differential operators defined in (\ref{complex sequence}).  On a
two dimensional Riemannian manifold, the Atiyah-Singer theorem reads
\begin{equation}
\chi(\Sigma_{g})=\sum_{r=0}^{2}(-1)^{r}h_{r}(\Sigma_{g}),
\end{equation}
where $\chi(\Sigma_{g})$ is the \textit{Euler characteristic} and
$h_{r}(\Sigma_{g})$ corresponds to the dimension of the Hodge groups
$H^{r}(\Sigma_{g}\otimes \ad P)$\cite{19}. It turns out that
$h_{0}=0$, since $h_{0}$ comes from a cohomology group of dimension
zero, more precisely it is the dimension of the space of sections of
$\ad P$ which are covariantly constant\cite{21}; the case of
$h_{2}$ is more subtle, and it is possible to pick the bundle
$\Sigma_{g}\otimes \ad P$ carefully ensuring that there are no
reducible connections \cite{23}; using this assumption
$h_{2}=0$ turning this surface into a \textit{bona fide} manifold.
Since $\chi(\Sigma_{g})=2-2g$ is the \textit{Euler characteristic}
of a Riemann surface of genus $g$, the dimension of the moduli space
for $g>0$ and $G$ simple is given by
\begin{equation}\label{dim2}
\dim \mathcal{M}_{F}(\Sigma_{g},G)=b_{1}(\Sigma_{g})=(2g-2)\dim G,
\end{equation}
where it has been used differential operators instead homotopy
groups. When the manifold is the two sphere, $g=0$, and
$\mathcal{M}_{F}(\Sigma_{g},G)$ is one point, this means that up to
gauge equivalence the only flat connection is the trivial
connection. For the torus the situation changes somewhat, as in the
homotopy approach the generators commute; this property is reflected
on the characteristic classes \cite{21} resulting
$\dim\mathcal{M}_{F}(\Sigma_{g},G)=2 \rank G$. In the case of
$U(1)$ as everything must commute then we have
$\dim\mathcal{M}_{F}(\Sigma_{g},G)=2g$.  Eventhough this approach
could not seem natural, it is possible to avoid some difficulties to
build loops on higher dimensional spaces with complex topologies. We
will use this approach in the next section for the case of four
dimensional $BF$ theories.

.\\

\noindent {\uno V. BF moduli spaces in four dimensions}
\vspace{.5em}

Since one of the equations of motion of BF theories is the vanishing
of the curvature (\ref{eq:BF}), it is possible to use some
properties of  instanton and anti-instanton complexes as the
condition  $F_{A}=0$ fullfills the usual self-dual and
anti-self-dual condition simultaneously, \textit{i.e.} the
connections that generate zero curvature are at the same time
self-dual and anti-self-dual connections. If we think the BF
equations as a non-linear generalization of Hodge theory such as in
the case of Yang-Mills equations, they are not elliptic as they
stand. The reason for this is that they possess a symmetry group;
from a physical point of view, this symmetry is the invariance under
gauge transformations; hence to obtain an elliptic problem we have
to choose a gauge. As we shall see in the next section, the gauge
transformations corresponding to a BF theory are given by
diffeomorphisms, in order to fix the gauge it is possible to make
use of the action (\ref{5}) which is not diffeomorphism invariant
and then take the limit when $g$ goes to zero.

\noindent Let $A$ a flat connection; when gauge equivalence is taken
properly into account, the space of such $A$ forms a finite
dimensional space $\mathcal{M}$ which we call the flat connection
moduli space. This space should be viewed as a finite dimensional
subspace of the infinite dimensional configuration space
$\mathcal{A/G}$, being $\mathcal{A}$ the space of all connections
and $\mathcal{G}$ the group of gauge transformations. To obtain a
\textit{good} moduli space we have to cut down, like the instantons
case, both the configuration space and the group defined on the
bundle, and to restrict $\mathcal{A/G}$ to the subspace of
irreducible connections\cite{21}. The reason for this is that one
needs to regularize the model to avoid problems with reducible
connections. Then there are some gauge transformations that act
trivially on the connections. This mean that $\mathcal{A/G}$ is not
in general a manifold as the quotienting out by the gauge group.
Generally the connections are irreducible, and there will be
isolated reducible connections; therefore $\mathcal{A/G}$ is then at
least an orbifold. Reducible connections are a source of great
difficulty in making sense of topological field theories in general.
The problem is that at reducible connections, path integrals related
to partitions functions diverge. Hence our new configuration space
is therefore the quotient
\begin{equation}
\mathcal{M}=\mathcal{A}^{irred}/\mathcal{G}.
\end{equation}

\noindent Our next task is to find the dimension of $\mathcal{M}$.
We employ a similar idea to that used on a Riemann surface in
section IV. The main idea is to work infinitesimally, by which we
mean to work with the tangent space to $\mathcal{M}$. The advantage
of doing this is that the dimension of the tangent space can be
calculated using the Atiyah-Singer theorem. Let $A+ta$ be a one
parameter family of smooth connections. Hence, by construction, $a$
is tangent to this family, so that
\begin{equation}
a\in T_{A}\mathcal{A}.
\end{equation}
We wish to obtain from $a$, the tangent space of the moduli space.
Using $[A]$ to denote the point of the moduli space to which $A$
belongs, then we obtain an element of $T_{[A]}\mathcal{M}$. Firstly
we must request this family to be flat, and secondly we must project
out those $a$'s which correspond to gauge directions, \textit{i.e.}
those $a$ which belong to the tangent space in the orbit-gauge
directions $T_{\mathcal{G}\cdot A}\mathcal{A}$. To achieve our first
condition from the equation of motion $F=0$ we see that
\begin{equation}
F(A+ta)=F(A)+td_{A}a+t^{2}a\wedge a,
\end{equation}
then, working infinitesimally $a$ satisfies $d_{A}a=0$. To achieve
our second goal we must identify those $a$ which differ by an
element of $T_{\mathcal{G}\cdot A}\mathcal{A}$. But, since
$T_{\mathcal{G}\cdot A}\mathcal{A}\simeq \Imag d_{A}$, it means
taking those $A$'s gauge equivalent; the two requirements are
satisfied if
\begin{equation} a\in \frac{\ker
d_{A}}{\Imag d_{A}},
\end{equation}
this has a cohomological interpretation which we now exploit using
the index theorem. The Lie algebra valued 1-forms $a$ are sections
of the bundle $\ad P\otimes \Lambda^{1}T^{*}M$, where the first
factor ensures that $a$ has the correct behavior under gauge
transformations, namely $a\mapsto g^{-1}ag$; the second factor is
simply because it is a 1-form. Let $\Omega^{i}(M,\ad P)$ the spaces
of sections where $\Omega^{i}(M,\ad P)=\Gamma(M,\ad P\otimes
\Lambda^{i}T^{*}M)$; the natural elliptic complex to use for our
index calculation is

\begin{equation}\label{complex}
0\stackrel{i}{\longrightarrow}\Omega^{0}(M,\ad
P)\stackrel{d_{A}}{\longrightarrow}\Omega^{1}(M,\ad
P)\stackrel{d_{A}}{\longrightarrow}\Omega^{2}(M,\ad
P)\stackrel{\pi_{+} \pi_{-}}{\longrightarrow}0.
\end{equation}
\noindent Where $\pi_{+},\; \pi_{-}$ are the operators which project
a two-form onto its self-dual part and onto its anti-self-dual part
respectively. This sequence of differential operators is defined as
in (\ref{complex sequence}) in order to calculate the moduli space
using the Atiyah-Singer theorem. It is a complex in the sense that
$d_{A}\circ d_{A}=0$. The cohomology data for this complex are

\begin{equation}
\begin{split}
H^{0}(E)& =\ker d_{A}^{(0)},\;\;\;\;\;\;\;\;\;\;\; \dim H^{0}(E)=h_{0},\\
H^{1}(E)& =\frac{\ker d_{A}}{\Imag d_{A}^{(0)}},\;\;\;\;\;\;\; \dim H^{1}(E)=h_{1},\\
H^{2}(E)& =\frac{\ker \pi_{+}\pi_{-}}{\Imag d_{A}},\;\;\;\;\;\;\;
\dim H^{2}(E)=h_{2},\\
\end{split}
\end{equation}
where $d_{A}^{(0)}$ denotes the exterior covariant derivative acting
on $\ad P\otimes \bigwedge^{0}T^{*}M$. Only one of these dimensions
corresponds to our moduli space calculation, this being $h_{1}$, in
other words, we wish to compute
\begin{equation}
h_{1}=\dim T_{[A]}\mathcal{M}=\dim \mathcal{M}.
\end{equation}
However, the index of the complex is the alternating sum
\begin{equation}
h_{0}-h_{1}+h_{2}.
\end{equation}
Nevertheless, it turns out that $h_{0}=0$ since $h_{0}$ comes from a
cohomology group of dimension zero, i.e. the dimension of space of
sections which are covariantly constant. Similary $h_{2}=0$, which
requires the use of a vanishing theorem. This is done by a
Bochner-Weitzenbech technique \cite{20} used in the case of
self-dual and anti-self-dual connections. This means that because of
a two-form can be expressed as a combination of its self-dual and
its anti-self-dual part respectively, both terms are mapped to zero
assuming that each term corresponds to the instanton and the
anti-instanton term in the Yang-Mills complex, in fact flat
connections satisfy both conditions; as we shall see this assumption
will restrict our manifold. The associated Laplacian
\begin{equation}
\Delta_{A}=d_{A}(d_{A})^{\dag}+(\pi_{+}\pi_{-})(\pi_{+}\pi_{-})^{\dag},
\end{equation}
is positive definite and hence has no kernel; the second term on the
right hand side is zero due to flatness condition on the curvature;
computing the remaining term in local coordinates shows that
\begin{equation}
\Delta_{A}=\frac{1}{2}d_{A}(d_{A})^{\dag}+\frac{R}{6}-W_{-}-W_{+},
\end{equation}
where $R$ is the scalar curvature of $M$ and $W_{+}$,$W_{-}$ the
self-dual part and the anti-self-dual part of its Weyl tensor.
Positivity will result if we assume $W_{+}$ and $W_{-}$ are zero;
that is, $M$ is known as a conformally flat manifold. The case when
$h_{2}\neq 0$ can be obtained considering corrections to the
associated Laplacian.    Since $h_{0}=0$ y $h_{2}=0$ we have
\begin{equation}
\Index=-h_{1}=-\dim \mathcal{M}.
\end{equation}
After complexification we can use our index formula (\ref{index})
\begin{equation}\label{4.21}
\Index=(-1)^{n(\frac{n}{2})}\int_{M}\frac{\ch(\sum_{p}(-1)^{p}[E^{p}])}{\e(M)}\cdot
\td(T(M_{C}))[M],
\end{equation}
in the present case $n=4$, and the $E^{p}$ are given by
\begin{align*}
E^{0}&=\ad_{C} P\otimes \Lambda^{0}T^{*}M_{C},& E^{1}&=\ad_{C}
P\otimes \Lambda^{1}T^{*}M_{C},& E^{2}&=\ad_{C} P\otimes
\Lambda^{0}T^{*}M_{C},
\end{align*}
with $\ad_{C} P$ the complexification of the adjoint bundle $\ad P$.
Now using the decomposition theorem of fiber bundles \cite{24,25}
and the multiplicative property of Todd classes, we have
\begin{equation}
T(M)_{C}=L_{1}\oplus \overline{L_{1}}\oplus L_{2}\oplus
\overline{L_{2}},
\end{equation}
then
\begin{equation}
\td(T(M_{C}))=(\frac{x_{1}}{1-\exp[-x_{1}]})(\frac{-x_{1}}{1-\exp[x_{1}]})(\frac{x_{2}}{1-\exp[-x_{2}]})(\frac{-x_{2}}{1-\exp[x_{2}]}),
\end{equation}
where $x_{1}$ and $x_{2}$ are two forms proportional to independent
eigenvalues of the curvature 2-form. In the same way using the
properties of the Euler class we have $\e(T(M))=x_{1}x_{2}$. To deal
with the rest of the formula we need to know
$\ch(E^{0}-E^{1}+E^{2})$, then using the properties of the Chern
character we have that
$\ch(E^{0}-E^{1}+E^{2})=\ch(\ad_{C}P)\ch(\Lambda^{0}T^{*}(M)_{C}-\Lambda^{1}T^{*}(M)_{C}+\Lambda^{2}T^{*}(M)_{C})$.
From the splitting principle \cite{25} we obtain that
\begin{equation}
\begin{split}
\ch(\Lambda^{0}T^{*}(M)_{C})& =1,\\
\ch(\Lambda^{1}T^{*}(M)_{C})&
=e^{x_{1}}+e^{-x_{1}}+e^{x_{2}}+e^{-x_{2}},\\
\ch(\Lambda^{2}T^{*}(M)_{C})&
=2+e^{x_{1}+x_{2}}+e^{x_{1}-x_{2}}+e^{-x_{1}+x_{2}}+e^{-x_{1}-x_{2}}.
\end{split}
\end{equation}
Finally replacing on the index formula (\ref{index}) we have
\begin{equation*}
\begin{split}
\Index (E)&
=\int_{M}[1-(e^{x_{1}}+e^{-x_{1}}+e^{x_{2}}+e^{-x_{2}})+(2+e^{x_{1}+x_{2}}+e^{x_{1}-x_{2}}+e^{-x_{1}+x_{2}}+e^{-x_{1}-x_{2}})]\\
& \frac{1}{x_{1}x_{2}} \cdot
(\frac{x_{1}}{1-\exp[-x_{1}]})(\frac{-x_{1}}{1-\exp[x_{1}]})(\frac{x_{2}}{1-\exp[-x_{2}]})(\frac{-x_{2}}{1-\exp[x_{2}]})\ch(\ad_{C}P).
\end{split}
\end{equation*}
Since the manifold is four-dimensional, no terms higher than 4-form
will appear in the characteristic classes, then developing the
polynomials we have

\begin{equation}\label{polynomials}
\Index
(E)=\int_{M}\ch(\ad_{C}P)(\frac{3}{x_{1}x_{2}}+ \frac{x_{1}}{x_{2}} + \frac{x_{2}}{x_{1}} + x_{1}x_{2}) [1 - \frac{1}{12} (x_{1}^{2} + x_{2}^{2})],
\end{equation}

 \noindent In four dimensions the Chern character takes the form
\begin{equation}
\ch(\ad_{C}P)=\rank(\ad_{C}P)+c_{1}(\ad_{C}P)+\frac{1}{2}(c_{1}^{2}(\ad_{C}P)-2c_{2}(\ad_{C}P)),
\end{equation}
however, $\ad_{C}P$ is the complexification of a real bundle, then
it is self-conjugated and has only even dimensional Chern classes;
also it is clear that $\rank(\ad_{C}P)=\dim G$. Finally we can
employ the properties of the Pontrjagin classes to write
$p_{1}(\ad_{C}P)=-2c_{2}(\ad_{C}P)$. This gives the result that
\begin{equation}\label{chernrank}
\ch(\ad_{C}P)=\dim G+\frac{1}{2}p_{1}(\ad_{C}P).
\end{equation}
Replacing (\ref{chernrank}) into (\ref{polynomials}) we have
\begin{equation} \label{Indiceintegral}
\Index (E) =\int_{M} (\dim G + \frac{1}{2} p_{1} (\ad_{C}P)) (\frac{3}{x_{1}x_{2}}+ \frac{x_{1}}{x_{2}} + \frac{x_{2}}{x_{1}} + x_{1}x_{2}) [1 - \frac{1}{12} (x_{1}^{2} + x_{2}^{2})];
\end{equation}
however, the singular terms $\frac{3}{x_{1}x_{2}}+ \frac{x_{1}}{x_{2}} + \frac{x_{2}}{x_{1}}$ require evidently a {\it regularization} in order to get regular polynomials. This is achieved considering that
\begin{equation}
\frac{3}{x_{1}x_{2}}+ \frac{x_{1}}{x_{2}} + \frac{x_{2}}{x_{1}} = \lim_{\xi\rightarrow 0} [\frac{3}{(x_{1}+\xi)(x_{2}+\xi)}+ \frac{x_{1}+\xi}{x_{2}+\xi} + \frac{x_{2}+\xi}{x_{1}+\xi}];\nonumber
\end{equation}
and making the expansion about the zero of the right-hand-side expression keeping only polynomials of order four we have an expression depending on
the Euler class $x_{1}x_{2}=e(M)$, and the Pontrjagin class
$x_{1}^{2}+x_{2}^{2}=p_{1}(M)$,
\begin{equation}
\frac{3}{x_{1}x_{2}}+ \frac{x_{1}}{x_{2}} + \frac{x_{2}}{x_{1}} = \lim_{\xi\rightarrow 0} \frac{1}{\xi^{4}} [ (3+\xi^{2})(x_{1}^{2}+x_{2}^{2}) + (3-2\xi^{2}) x_{1}x_{2} + \xi^{2} (3+2\xi^{2})];\nonumber
\end{equation}
{\it regularization} requires then the integration of the singular terms through $\lim_{\xi\rightarrow 0}\int \lambda \xi^{4}$ (singular terms), where $\lambda$ is a global factor to be determined:
\[
\Index (E) = \int_{M} \dim G [(3\lambda+1)x_{1}x_{2} + 3\lambda (x_{1}^{2}+x_{2}^{2})],
\]
where we have considered that $p_{1}(\ad_{C}P)$ is proportional to a 4-form. Let for example $E$ be a $SU(2)$-bundle, then $E$ carries the fundamental two dimensional representation of $SU(2)$ and, if we
make the tensor product of $E$ with itself, there is a natural
decomposition of this tensor product bundle into three-dimensional
and one-dimensional representations. However, because $E\otimes E$
is quadratic in $E$, it is clear that the elements of the
fundamental representation are both mapped onto the same element in
the tensor product, thus the bundle $\ad_{C}P$ is the three
dimensional part of the tensor product $E\otimes E$ \cite{25}.
Decomposing $E\otimes E$ into the sum of a symmetric  and an
anti-symmetric part
\begin{equation}\label{decomposition EE}
E\otimes E=S^{2}E\oplus \Lambda^{2}E,
\end{equation}
then applying the properties of the Chern character to
(\ref{decomposition EE}) gives
\begin{equation}\label{Chern EE}
\ch(E)\ch(E)=\ch(\ad_{C}P)+\ch(\Lambda^{2}E),
\end{equation}
if we expand both sides using the properties of Chern character we
get
\begin{equation}
(2+c_{1}(E)+\frac{1}{2}(c_{1}^{2}(E)))^{2}=3+\frac{1}{2}p_{1}(\ad_{C}P)+1+c_{1}(E).
\end{equation}
But on $M$ we need  only keep polynomials of dimension four, so that
\begin{equation}\label{p1ad}
\begin{split}
4-4c_{2}(E)& =4+\frac{1}{2}p_{1}(\ad_{C}P)\\
\Rightarrow p_{1}(\ad_{C}P)& =-8c_{2}(E),
\end{split}
\end{equation}
which corresponds essentially a 4-form.
As we expected our Index will be written in terms of topological invariants
that describe global properties of the $SU(N)$-bundle and of the
background manifold $M$. To see this, according to the Gauss-Bonnet
theorem \cite{24} and to the Hirzebruch signature theorem \cite{26}
the {\it virtual} dimension of the moduli space for a BF theory on a $SU(N)$-bundle is given
finally by
\begin{equation}
h_{1} =- \Index (E) = -\dim G [(3\lambda+1)\chi + 9\lambda |\tau|];
\end{equation}
where $SU(N)$ is the structure group of the bundle that in
the fundamental representation has $dim G = N^{2}-1$. This index represents on the one hand the dimension of
flat connections, and on the other hand the dimension of the intersection of the spaces of connections that
generate 4-instantons and 4-anti-instantons simultaneously.

Now we need to fix $\lambda$ in order to obtain values of $h_{1}$ physical and geometrically admissible as moduli space dimension. One may to try with different values of $\lambda$, but the algebraic structure of the above expression and the fact that in general $\chi \geq |\tau|$ (see the tables below), lead to an expression essentially of the form $\alpha(\lambda) (m|\tau| - \chi)$, with $m$ rational, and $\alpha$ a constant depending on $\lambda$; a direct comparison leads to $3\lambda+1=\alpha$, and $-9\lambda=\alpha m$, which allows to obtain an expression in terms of $m$:
\begin{equation}
h_{1} = \dim G \frac{3}{3+m} (m|\tau| -\chi);
\end{equation}
reducing the problem of fixing $\lambda$ to choose an appropriate rational number $m$. If $m<0$, positivity of $h_{1}$ will require $m<-3$; however, in the case of $S^{4}$ as base manifold this condition will lead to a moduli space dimension of flat connections bigger than the corresponding to instanton or anti-instantons, which is inadmissible since flat connections can be viewed as the intersection of the space of those field configurations; hence we can consider as first restriction $m>-3$. More specifically if we consider that the dimension of the moduli space for $SU(2)$-instantons on $S^{4}$ with instantonic number $k=0$ is 5, then the restriction is $\lambda > -11/18$, and considering that $\lambda =\frac{m}{9+3m}$ we obtain consistently the restriction $m>-3$, at least for the case $S^{4}$. However, this restriction on  $m$ leads to $h_{1}<0$, and the moduli space will be considered empty.
Using the equation (\ref{dim2}) we can observe the case of a two-dimensional sphere $S^{2}$, and that it is true also for a four-dimensional
sphere $S^{4}$, both have as dimensional moduli spaces just one point, this means that up to gauge transformations the only flat
connection is the trivial connection, property observed in homotopy theory and in other calculus with different complexes
\cite{27,28,29}.

For most of base manifolds, the cases with integers $m=-2, -1, 0, 1$ are ruled out due to yield non-positive $h_{1}$ and the corresponding moduli space is considered empty; additionally $m=2$ is also ruled out due to yields positive but non-integer $h_{1}$; but this last value may make sense only for $SU(4)$ since $\dim [SU(4)]=15$, which is divisible by 5. However, there will exist fractional values of $m$ leading to integer $h_{1}$ for arbitrary gauge symmetry group as we shall se below in the figures.

Continuing with integer values of $m$ that yield admissible values for $h_{1}$, let us see now the case $m=3$, and hence $h_{1}=\frac{1}{2} \dim G (3|\tau|-\chi)$; in the following table we display explicitly the values of $h_{1}$ for different background four-manifolds. The symbol $\emptyset$ denotes a negative virtual dimension, and will be understood as a empty
moduli space; it is different of course from a zero dimensional moduli space.

\begin{tabular}[t]{|c| c| c|c|}
\hline \multicolumn{4}{|c|}{Table 1: Characteristic numbers with $m$=3}\\
\hline
  {}& $\chi$ & $\tau$ & $h_{1}=\frac{1}{2} \dim G (3|\tau|-\chi)$\\
  \hline
  $S^{4}$ & 2 & 0 & $\emptyset$ \\
 \hline
 $CP_{2}$ & 3 & 1 & $0$\\
 \hline
 $S^{2}\times\Sigma_{g} $ & 4(1-$g$) & 0 & $\emptyset$ \textit{for} \ $g=0$; \\
 {} & {} & {} &$ 2\dim G(g-1)$ \textit{for} \ $g\geq 1$;\\
 \hline
$K3$& 24& -16& $12\dim G$\\
\hline
 $K3_{Z_{2}}$& 12& -8& $ 6\dim G$\\
 \hline
 $K3_{Z_{2}\otimes Z_{2}}$& 6& -4& $3\dim G$\\
 \hline
 $E(n)$ & 12$n$ & -8$n$ & $6n\dim G$ \\
 \hline
 $S_{d}$ & $d(6-d+d^{2})$ & $\frac{1}{3}(4-d^{2})d$ & $0 \ for \ d=1$;\\
 {} & {} & {} &$\emptyset$ \textit{for} \ $d=2$;\\
 {} & {} & {} & $\dim G \ d(2d-5) \ for \ d>2$ \\
 \hline
\end{tabular}
\\ \\

\noindent $S^{2} \times \Sigma_{g}$ represent product manifolds of $S^{2}$ with Riemann surfaces of genus $g$. Note that for $g=0$, we have $S^{2} \times S^{2}$, and $h_{1}$ is negative and we shall consider it empty; for $g\geq 1$, $h_{1}$ is a non-negative number. $Sd$ represent hypersurfaces of degree $d$ in $CP(3)$ associated with the homogeneous polynomials $\sum^{4}_{i=1} z^{d}_{i}=0$; for example, $S_{4}$ represents the $K3$ surface.
 $E(n)$ represent the so called elliptic surfaces, which can be viewed also as elliptic fibrations where the fibers correspond to elliptic curves; these simply connected four-dimensional manifolds are labeled by a non-negative integer $n$ . For example $E(2)$ reduces in particular to $K3$.

The cases with $m=4,5$ are ruled out by similarity with the case $m=2$. The cases $m=6,15$ are meaningful with $h_{1} = \frac{1}{3}\dim G (6|\tau|-\chi)$, and $h_{1}= \frac{1}{6}\dim G (15|\tau|-\chi)$ respectively,
and in tables 2 and 3 we display the corresponding characteristic numbers. In the row corresponding to $S_{d}$, a hat $\widehat{}$ means that the number must be omitted from the sequence.

\begin{tabular}[t]{|c| c| c|}
\hline \multicolumn{3}{|c|}{Table 2: Characteristic numbers with $m$=6 and $h_{1}=\frac{1}{3}\dim G(6|\tau|-\chi)$}\\
\hline
  {}& $6|\tau|-\chi$ & $h_{1}$\\
  \hline
  $S^{4}$ & -2 & $\emptyset$\\
 \hline
 $CP_{2}$ & 3 & $\dim G$\\
 \hline
 $S^{2} \times \Sigma_{g}$ & $4(g-1)$ & $\frac{4}{3}(g-1) \dim G \ for \ SU(N) \ and \ g=3l+1, \ l=0,1,2,3 \ldots;$\\
 {} & {} & $4(g-1) \ for \ SU(2), \ and \ g=1,2,3,4 \ldots;$\\
 {} & {} & $20(g-1) \ for \ SU(4), \ and \ g=1,2,3,4 \ldots;$\\
 {} & {} & $\emptyset$ \textit{for} \ $g=0$; \\
 \hline
 $K3$& 3(24)&  $24\dim G$\\
 \hline
 $K3_{Z_{2}}$& 3(12)& $12\dim G$\\
 \hline
 $K3_{Z_{2}\otimes Z_{2}}$& 12& $4\dim G$\\
 \hline
 $E(n)$ & 3(12$n$) & $12n\dim G$ \\
 \hline
 $S_{d}$ &  $3 \ for \ d=1;$ & $\dim G;$\\
 {} & $-4 \ for \ d=2;$ &  $\emptyset$; \\
 {} & $d[d^{2}+4(d-3)]$ & $\frac{1}{3} \dim G d[d^{2}+4(d-3)] \ for \ SU(N) \ and \
 d=3,\widehat{4},5,6,\widehat{7},8,9,\widehat{10},11 \ldots;$\\
 {} & $for \ d>2;$ & $d[d^{2}+4(d-3)] \ for \ SU(2);$\\
 \hline
\end{tabular}
\\ \\

\begin{tabular}[t]{|c| c| c|}
\hline \multicolumn{3}{|c|}{Table 3: Characteristic numbers with $m$=15 and $h_{1}= \frac{1}{6}\dim G (15|\tau|-\chi)$}\\
\hline
  {}& $ 15|\tau|-\chi$ & $h_{1}$\\
  \hline
  $S^{4}$ & -2 & $\emptyset$ \\
 \hline
 $CP_{2}$ & 12 & $2\dim G$\\
 \hline
 $S^{2} \times \Sigma_{g}$ & $4(g-1)$ & $\frac{2}{3} \dim G(g-1) \ for \ SU(N) \ and \ g=3l+1, \ l=0,1,2,3 \ldots;$\\
 {} & {} & $2(g-1) \ for \ SU(2), \ and \ g=1,2,3,\ldots;$\\
 {} & {} & $\emptyset$ \textit{for} \ $g=0$;\\
 \hline
 $K3$& 9(24)&  $36\dim G$\\
 \hline
 $K3_{Z_{2}}$& 9(12)& $18\dim G$\\
 \hline
 $K3_{Z_{2}\otimes Z_{2}}$& 6(9) & $9\dim G$\\
 \hline
 $E(n)$ & 9(12)$n$ & $18n\dim G$ \\
 \hline
 $S_{d}$ &  $12 \ for \ d=1;$ & $2\dim G;$\\
 {} & $-4 \ for \ d=2;$ &  $\emptyset$ ; \\
 {} & $2d[2d(d+1)-13)]$ & $d[2d(d+1)-13] \ for \ SU(2);$\\
 {} & $for \ d>2;$ & $\frac{1}{3} \dim G d[2d(d+1)-13] \ for \ SU(N) \ and \ d=3,4,\widehat{5},6,7,\widehat{8},9,10,\widehat{11} \ldots;$\\
 \hline
\end{tabular}
\\ \\

Besides the four-dimensional sphere, there exist two cases with empty moduli space,
$S^{2} \times \Sigma_{0}= S^{2} \times S^{2}$ and $S_{2}$; however this fact is not fortuitous, since these last four-dimensional manifolds are diffeomorphic to each other $S^{2} \times S^{2}\simeq S_{2}$; consistently it is well known that there not exist instantons (nor anti-instantons) on $S^{2} \times S^{2}$, which is compatible with our interpretation
of flat connections lying in the intersection, in this case, of empty spaces.

\noindent Although in the tables the values of $h_{1}$ are displayed for fixed values of $m$, it is
necessary to have a general outlook of the behavior of $h_{1}$ without restrictions on $m$ for different base manifolds; hence
the tables correspond only to points in the figure 1 with both $h_{1}$ and $m$ positive integers;
these tables can be considered as a display zoom of points on the different curves.
Such points are in the region restricted for admissible values of $h_{1}$ leading to real
moduli space dimensions. However, from this global view it is impossible to look the asymptotic
behavior of $h_{1}$ as $m\rightarrow-3$, where there exist admissible integer values as moduli space dimension. A display zoom of this asymptotic region is given in the figure 2 for $CP_{2}$, showing the
generic behavior of the dimension for all base manifolds considered in figure 1;
in all cases there will be an infinite (but countable) number of values admissible as real dimensions.
Additionally it is possible to define invariants of differentiable
structures counting the number of points in zero dimensional moduli
spaces like in the case of monopole moduli spaces \cite{30};
in the present case all curves have a cross point with the axis $h_{1}=0$,
condition satisfied by the rational $m=\chi/\mid\tau\mid$, except of course the case of manifolds
with $\tau=0$, such as $S^{2} \times \Sigma_{g}$.\\
\begin{figure}[!hpb]
\centering
 \includegraphics[width=0.9\textwidth]{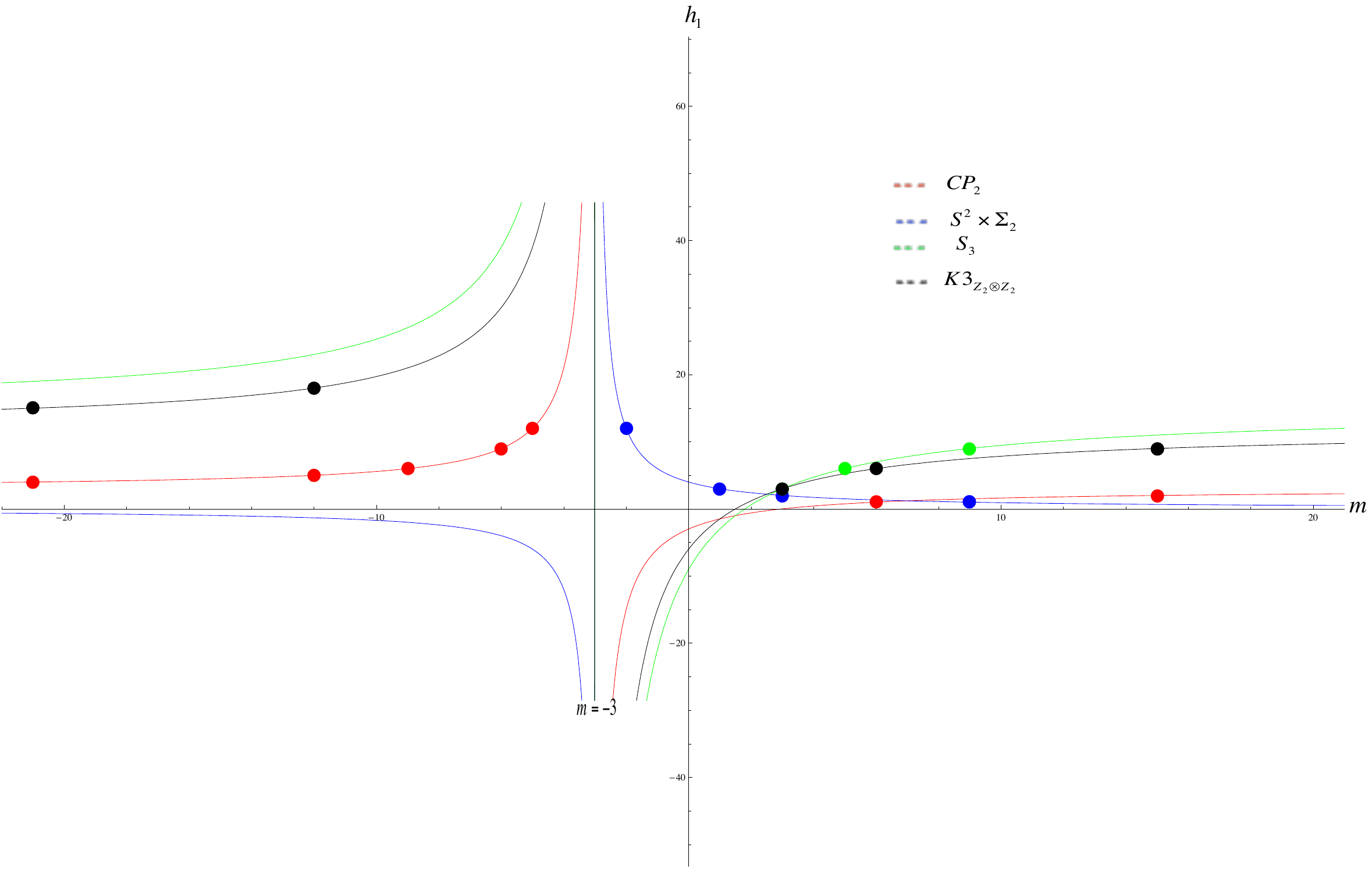}
\hfil
  \caption{The {\it virtual} dimension $h_1$ is represented for different base
manifolds using $[dimG]$ as unit. The continuous lines represent $h_1$ for  $m \in
R$, and the points represent non-negative
integers for $m$ rational. The vertical asymptote corresponds to $lim_{(m \rightarrow-3) } h_1= \pm \infty$; there exists a horizontal asymptote for each
manifold, and corresponds to $lim_{(m \rightarrow \pm \infty) } h_1= 3 \mid \tau\mid $, separating
two disconnected parts of the curves. On this figure the fist criterion
for obtaining admissible values is $h_{1}$ $\geq$ 0;  although for
the virtual dimension $m \in R$, the integer values of $h_1$ are
always contained in a finite range  $m \in  \{ a,b \}$
,with $a$ and $b$ integers, with an infinite and countable  number of values of  $h_1$  (see figure 2).
For example, the range for $CP_{2}$ is  $\{-21, 15\}$.
The curves for $S^{2}\times \Sigma_{2}$, $K3_{Z_{2}\otimes Z_{2}}$, and $S_{2}$ represent the global behavior of the families $S^{2}\times \Sigma_{g}$, $K3$, and $S_{d}$ respectively.}
\label{figure 1}
\end{figure}
\begin{figure}
  \centering
\includegraphics[width=0.9\textwidth]{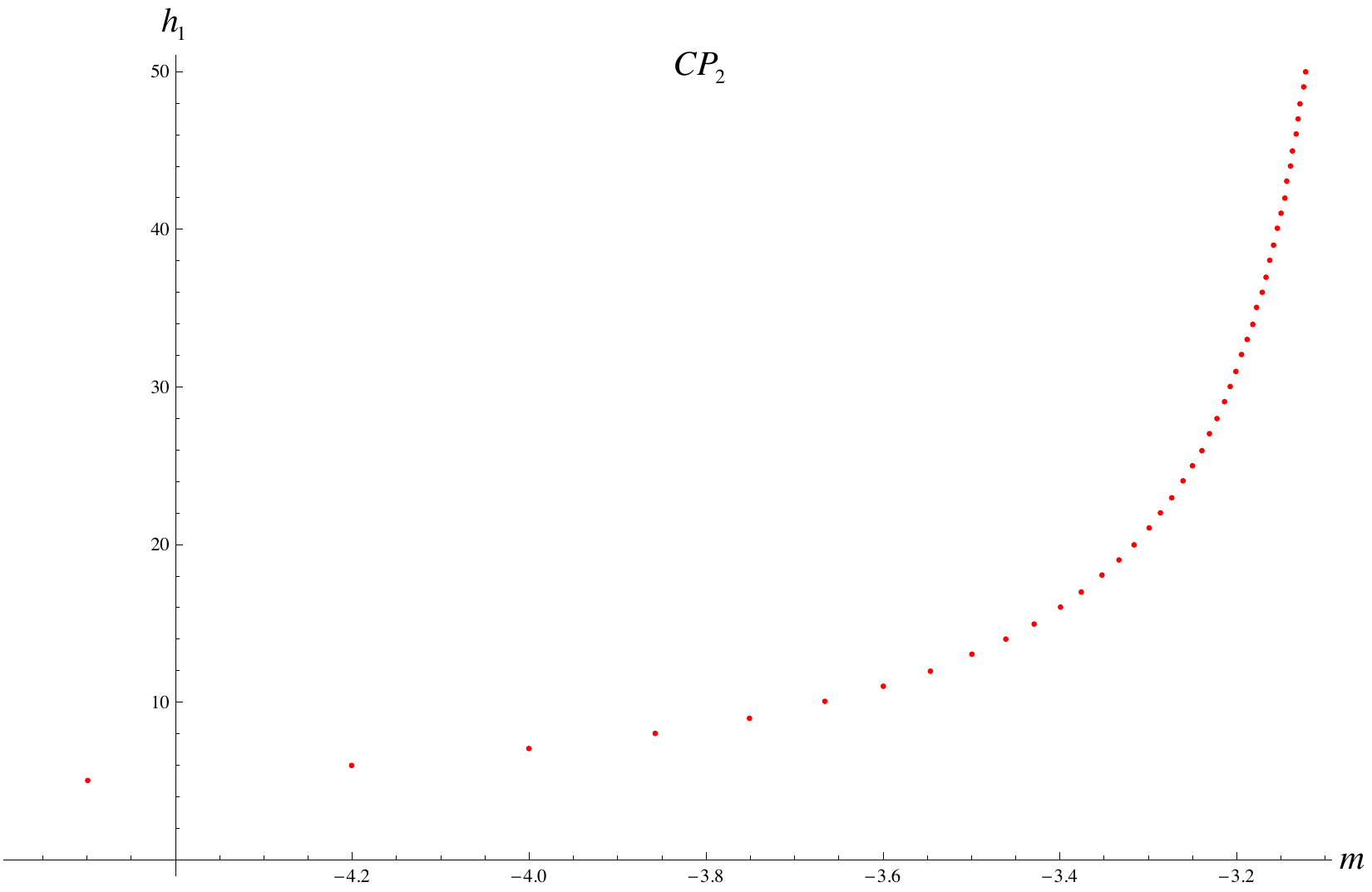}
     \caption{The limiting behavior of $h_1$ as $m\rightarrow -3$; $m=-3$ is an accumulation point for non-negative integers of $h_1$
which are represented by the points in the figure; therefore, the
physically admissible values correspond to a set infinite and countable.This figure has been obtained by mean of the below {\it fifty points}-sequence of the form $\{m,h_{1}\}$ with $m$ rational and $h_{1}$ integer, generated from the inverse function $m=m(h_{1})$ of the Eq.\ (41) for $CP_{2}$, and evaluated for the integers $h_{1}=1,2,..,50,$ in $[\frac{1}{3}dimG]$ units; one can obtain other {\it n-points} sequences for arbitrary $n$.}
  \label{figure 2}
\end{figure}
\newpage
\begin{eqnarray}
\{\infty,1\}&,& \{-9, 2\}, \quad \{-6, 3\}, \quad \{-5, 4\},  \quad  \{-\frac{9}{2}, 5\},  \quad  \{-\frac{21}{5}, 6\},  \quad  \{-4, 7\},  \quad  \{- \frac{27}{7}, 8\},  \quad  \{-\frac{15}{4}, 9\},  \nonumber \\
 \{- \frac{11}{3}, 10 \}&,&  \{-\frac{18}{5}, 11\}, \quad \{- \frac{39}{11}, 12\},  \quad  \{- \frac{7}{2}, 13\},  \quad \{- \frac{45}{3}, 14\},  \quad  \{- \frac{24}{7}, 15\},  \quad  \{- \frac{17}{5}, 16\},   \nonumber \\
\{- \frac{27}{8}, 17\}&,&  \{- \frac{57}{17}, 18\},  \quad  \{- \frac{10}{3}, 19\}, \quad  \{- \frac{63}{19}, 20\},  \quad \{- \frac{33}{10}, 21\}, \quad  \{- \frac{23}{7}, 22\}, \quad  \{- \frac{36}{11}, 23\},  \nonumber \\
\{- \frac{75}{23}, 24\}&,&  \{-\frac{13}{4}, 25\}, \quad  \{- \frac{81}{25}, 26\},  \quad  \{-\frac{42}{13}, 27\},  \quad  \{- \frac{29}{9}, 28\}, \quad  \{- \frac{45}{14}, 29\}, \quad  \{- \frac{93}{29}, 30\},   \nonumber \\
\{- \frac{16}{5}, 31\} &,& \{-\frac{99}{31}, 32\}, \quad \{- \frac{51}{16}, 33\},  \quad \{- \frac{35}{11}, 34\},  \quad \{- \frac{54}{17}, 35\},  \quad \{- \frac{111}{35}, 36\},  \quad \{- \frac{19}{6}, 37\},    \nonumber \\
 \{- \frac{117}{37}, 38\}&,& \{-\frac{60}{19}, 39\},   \quad \{- \frac{41}{13}, 40\},  \quad \{- \frac{63}{20},  41\},  \quad \{- \frac{129}{41}, 42\},   \quad\{- \frac{22}{27}, 43\},   \quad \{- \frac{135}{43}, 44\},    \nonumber \\
\{- \frac{69}{22}, 45\}&,&  \{- \frac{47}{15}, 46\}, \quad \{- \frac{72}{23}, 47\},  \quad \{- \frac{147}{47}, 48\},  \quad \{- \frac{25}{8}, 49\},  \quad \{- \frac{153}{49}, 50\}.
\end{eqnarray}
 As discussed previously, the expression for the {\it virtual} dimension makes sense for $S^{4}$ only under the restriction $m>-3$; if this restriction
is imposed as universal criterion on the other base manifolds, then only the region defined by $m>-3$ and $h_{1}\geq 0$ will contain the admissible values as real dimensions. Under these conditions the only case with a infinite and countable number of values will be $S^{2} \times \Sigma_{g}$, being the other cases a finite number of isolated points in the permitted region (the branches of the curves to the left of the asymptote $m=-3$ will be not admissible). However, it is not the unique criterion, since that the expression for the
{\it virtual} dimension has been obtained under the only assumption of a 4-dimensional background compact manifold, and it is reasonable consider that
will admit independent restrictions for obtaining real dimensions; hence under this new criterion the restriction valid for the four-dimensional sphere will not affect to the curves corresponding to other manifolds, being the restriction $h_{1}\geq 0$ the only universal criterion (and the branches of the curves eliminated under the first criterion will be now recovered). Anyway, the results of the tables 1,2, and 3 are valid under both criteria (they were constructed deliberately taken into the account these considerations).\\
It is mandatory to try find possible physical or geometrical explanations on the appearance of a rational $m$ characterizing
the moduli space dimensions for flat connections; a possibility is as follows, and it is connected with the question formulated in the introduction
on a possible relationship between 4-dimensional YM instanton moduli space and the corresponding one to 4-dimensional BF field theory, interpreted here as 4-dimensional flat
connections lying in the intersection of the instantons and anti-instantons spaces. The later are characterized as well known by the instantonic number $k$
(infinite and countable), and defined in terms of the squared norm of the self-dual or anti-self-dual curvature respectively. Instead a rational $m$
(infinite and countable) appears in the case at hand; therefore, the relationship mentioned may be through a possible correlation between the numbers $k$, and $m$: for
different $m$'s, we shall have in general different intersections of the instantons and anti-instantons spaces, and such intersection spaces
will be labeled with a $m$ through the correlation $m=m(k)$ (finally both $k$ and $m$ are infinite and countable). The seeking for
this correlation (if any) will be the subject of forthcoming communications.\\

\noindent {\uno VI. Dirac's canonical analysis} \vspace{.5em}\\
In this section, we shall develop Dirac's canonical analysis for a
four-dimensional modified BF theory, reproducing on shell   topological YM
theory. We will see  later  that the theory studied in this section  shares the same moduli space with both  BF theory and
the BF-YM theory in the limit when the gauge coupling goes to zero,
and additionally  preserves the same gauge symmetries of BF theory at Hamiltonian level.  The
Hamiltonian framework developed in this section, will allow us to
understand the principal symmetries of the theory as well as its
constraints, the extended Hamiltonian and the gauge transformations.
The theory under study will  depend on a connection valued in the
Lie algebra of $SU(N)$  \cite{31,32}.\\
Our starting  point is  the following action
\begin{equation}\label{action BF}
S[A,B]=\int_{M} Tr \left(i B\wedge F(A)+\frac{g^2}{4}B\wedge
B\right),
\end{equation}
\noindent where
$F_{\mu\nu}^{I}=\partial_{\mu}A_{\nu}^{I}-\partial_{\nu}A_{\mu}^{I}+f^{IJK}A_{\mu}^{J}A_{\nu}^{K}$
is the curvature of the connection 1-form $A_{\mu}^{I}dx^{\mu}$;
being $f^{IJK}$ the structure constants of the Lie algebra $SU(N)$
and $B_{\alpha \beta}^{I}$ is a set of $6(N^{2}-1)$ $SU(N)$
components of valued 2-forms. Here, $\mu, \nu=0,1,\ldots, 3$ are
spacetime indices, $x^{\mu}$ are the coordinates that label the
points of the 4-dimensional manifold $M$ and
$I,J,K=0,1,\ldots,N^{2}-1$, are the internal indices that can be
raised and lowered by the
Cartan-Killing metric given by the Lie algebra.\\
The action (\ref{action BF}) yields the next equations of motion
\begin{equation}\label{motioneq}
F^{I}(A)=i\frac{g^{2}}{2}B^{I}, \;\;\;\;\;\;\; DB^{I}=0,
\end{equation}
\noindent where $F^{I}$ satisfies Bianchi's identities $DF^{I}=0$.
By substituting the equations of motion (\ref{motioneq}) into
(\ref{action BF}) we obtain the topological YM theory. To perform
the Hamiltonian analysis, we shall consider that the manifold $M$
has a topology $\Sigma \times R$, where $\Sigma$ corresponds to a
Cauchy surface and $R$ represents an evolution parameter.  In this
manner, by making the $3+1$ decomposition  the action (\ref{action
BF}) takes the form
\begin{equation}\label{action BF2}
S[A,B]=\frac{1}{2}\int\int_{\Sigma}
d^{3}xdt\epsilon^{0ijk}\{i(\dot{A}_{k}^{I}-D_{k}A_{0}^{I})B_{ij}^{I}+B_{0i}^{I}(iF_{jk}^{I}+\frac{g^{2}}{2}B_{jk}^{I})\},
\end{equation}
\noindent where we are able to  identify the corresponding Lagrangian density
\begin{equation}
\mathcal{L}=\frac{1}{2}\eta^{ijk}\{i(\dot{A}_{k}^{I}-D_{k}A_{0}^{I})B_{ij}^{I}+B_{0i}^{I}(iF_{jk}^{I}+\frac{g^{2}}{2}B_{jk}^{I})\},
\end{equation}
\noindent here $\epsilon^{0ijk}\equiv\eta^{ijk}$, $i,j,k=1,2,3$ and
$\eta^{123}=1$. To carry out the Hamiltonian analysis, we will
consider as dynamical variables those with time derivatives
occurring in the action; an alternative procedure  can also be
considered in \cite{33}, where a pure Dirac's analysis of other
topological theories is performed. \\ Then,  the canonically
conjugate momenta $\Pi^{iI}$ to the $A_{i}^{I}$ are given
\begin{equation}\label{defmomenta}
\Pi^{i I}\equiv \frac{\delta \mathcal{L}}{\delta
\dot{A}_{i}^{I}}=\frac{i}{2}\eta^{ijk}B_{jk}^{I}.
\end{equation}
In this manner, by using the definition of the momenta  in the action
(\ref{action BF2}) we obtain
\begin{equation}\label{action BF3}
S[A_{i}^{I},\Pi^{i
I},B_{0i}^{I},A_{0}^{I}]=\int_{M}d^{4}x\left( \Pi^{iI}\dot{A}_{i}^{I}-A_{0}^{I}D_{k}\Pi^{k
I}-\frac{i}{2}\eta^{ijk}B_{0i}^{I}F_{jk}^{I}+i\frac{g^{2}}{2}\Pi^{i
I}B_{0i}^{I} \right).
\end{equation}
From the action (\ref{action BF3}) we can identify the non-vanishing
fundamental Poisson brackets for the theory
\begin{equation}
\{A_{i}^{I}(x^{0},\vec{x}),\Pi^{j
}_J(x^{0},\vec{y})\}=\delta_{i}^{j}\delta^{I}_{J}\delta^{3}(x-y),
\end{equation}
\noindent
and  the corresponding Hamiltonian of this theory  given by
\begin{equation}\label{hamiltonian}
H_{c}= \dot{A}_{i}^{I}\Pi^{iI}- \mathcal{L}=  \int d^{3}x \left(-A_{0}^{I}D_{k}\Pi^{k
I}-iB_{0i}^{I}\left(\frac{1}{2}\eta^{ijk}F_{jk}^{I}-\frac{g^{2}}{2}\Pi^{i
I}\right)\right).
\end{equation}
Calculating the variation of (\ref{action BF3}) with respect to
$A_{i}^{I}$, $\Pi^{iI}$  the equations of
motion read
\begin{equation}
\begin{split}
\delta A_{i}^{I}\; &:\; \eta^{ijk}D_{j}B_{0k}^{I}=0,\\
\delta \Pi^{iI}\; &:\;
D_{i}A_{0}^{I}=\dot{A}_{i}^{I}-i\frac{g^{2}}{2}B_{0i}^{I},
\end{split}
\end{equation}
\noindent and the variations respect to $B_{0i}^{I}$, and
$A_{0}^{I}$ yield the following $4(N^{2}-1)$ primary constraints
\begin{equation}\label{constraints}
\begin{split}
\phi^{I}\; &: \;D_{k}\Pi^{k I}\approx 0,\\
\phi^{i I}\;&:\;\frac{g^{2}}{2}\Pi^{i
I}-\frac{1}{2}\eta^{ijk}F_{jk}^{I}\approx 0.
\end{split}
\end{equation}
\noindent
As we can observe, the Hamiltonian (\ref{hamiltonian}) is
a linear combination of the constraints (\ref{constraints}) and
$A_{0}^{I}$, $B_{0i}^{I}$, both
correspond to Lagrange multipliers.\\
Now, we need to identify whether the theory presents secondary
constraints. From the temporal evolution of the constraints
(\ref{constraints}), we can observe that consistency demands that
there are no more
  constraints because
\begin{equation}\begin{split}
\dot{\phi}^{I}&=\{\phi^{I}(x),H_{c}\}= f^{IJK}\left[A_{0}^{J}\phi^{K}-\phi^{i J}B_{0i}^{K}\right] \approx 0,\\
\dot{\phi}^{iI}&=\{\phi^{iI}(x),H_{c}\}=f^{IJK}A_{0}^{J}\phi^{iK}\approx
0.
\end{split}
\end{equation}
With all constraints at hand,  we need to identify  which ones
correspond to first and second class. In order to do this, we need
to calculate the Poisson brackets between all the constraints, which
are given by
\begin{equation}\label{constraint-algebra}
\begin{split}
\{\phi^{I}(x),\phi^{J}(y)\}&=f^{IJK}\phi^{K}\delta^{3}(x-y),\\
\{\phi^{I}(x),\phi^{iJ}(y)\}&=f^{IJK}\phi^{iK}\delta^{3}(x-y),\\
\{\phi^{i I}(x),\phi^{j K}(y)\}&=0,
\end{split}
\end{equation}
\noindent thus, we  observe that the constraints are of first
class. Nevertheless, we can see  that the
$4(N^{2}-1)$ first class constraints given in (\ref{constraints})
are not all independents. The reason is because of Bianchi's
identity $DF^{I}=0$ implies
\begin{equation}
D_{i}\phi^{i I}=\phi^{I}.
\end{equation}
\noindent Thus, from the $3(N^{2}-1)$ first class constraints
$\phi^{i I}(x)$, we identify that
$[3(N^{2}-1)-(N^{2}-1)]=2(N^{2}-1)$ are independents. Therefore, we
are able to calculate the physical degrees of freedom as follows; we
have $6(N^{2}-1)$ canonical variables, $3(N^{2}-1)$ independent
first class constraints and there are not second class constraints.
With this information, we conclude that the action (\ref{action BF}) is devoid of physical  degrees of freedom; this scheme can be
generalized to other BF theories \cite{33,34}. As we can observe,
the action defined in (\ref{5}) and (\ref{action BF}) share a kind
of similarity,  but (\ref{action BF}) has the presence of the
Hodge-duality operation, and this  fact allows the theory   has
$2(N^{2}-1)$ degrees of freedom. Nevertheless,
the action given in (\ref{action BF}) has not the duality operator and the theory is devoid of physical degrees of freedom.   \\
The identification of the constraints allows us to construct  the
extended action which is given by
\begin{equation}\label{Eaction}
S_{E}[A_{i}^{I}, \Pi^{i I}, A_{0}^{I},
B_{0i}^{I}]= \int_{M}d^{4}x \Big\{ \dot{A_{i}^{I}}\Pi^{i
I}+A_{0}^{I}D_{i}\Pi^{i
I}+\frac{1}{2}\eta^{ijk}B_{0i}^{I}F_{jk}^{I}+\Pi^{i I}B_{0i}^{I}\Big\}.
\end{equation}
\noindent From (\ref{Eaction}) we can identify the extended
Hamiltonian
\begin{equation}
H_{E}=-A_{0}^{I}D_{i}\Pi^{i
I}-\frac{i}{2}B_{0i}^{I} \left(\eta^{ijk}F_{ij}^{I}-g^{2}\Pi^{i I} \right),
\label{eq57}
\end{equation}
\noindent that is a linear combination of first class constraints as
 expected. \noindent As well know, the equations of motion obtained
from the extended Hamiltonian in general are mathematically
different with the
Euler-Lagrange equations, but the difference is unphysical. \\
The equations of motion obtained from the extended action are
\begin{equation}
\begin{split}
\delta A_{0}^{I}\; &:\; D_{i}\Pi^{i I}=0,\\
\delta B_{0i}^{I}\; &: \;
\frac{1}{2}\epsilon^{0ijk}F_{jk}^{I}-\frac{g^{2}}{2}\Pi^{iI}=0,\\
\delta A_{i}^{I}\; &:\; \epsilon^{ijk}D_{j}B_{0k}^{I}=0,\\
\delta\Pi^{iI}\;
&:\;D_{i}A_{0}^{I}=\dot{A}_{i}^{I}-i\frac{g^{2}}{2}B_{0i}^{I}.
\end{split}
\end{equation}
Now we proceed computing the gauge transformations on the phase
space. To this aim, we need to use the first class constraints to
define the generator of gauge transformations as
\begin{equation}
G=\int_{\Sigma}\varepsilon^{I}\phi^{I}+\varepsilon_{i}^{I}\phi^{i
I},
\end{equation}
\noindent thus, we find the following gauge transformations on the
phase space
\begin{equation}\label{gaugetransformations}
\begin{split}
\delta_{0}A_{i}^{I}&=-D_{i}\varepsilon^{I}+\varepsilon_{i}^{I},\\
\delta_{0}\Pi^{i I}&= f^{IJK}\varepsilon^{J}\Pi^{i
K}+\epsilon^{0ijk}D_{j}\varepsilon_{k}^{I},\\
\delta_{0}A_{0}^{I}&=0,\\
\delta B_{0i}^{I}&=0.
\end{split}
\end{equation}
On the other hand,  we know that the $BF$ theory is diffeomorphisms
covariant,  and apparently that symmetry is not present in
(\ref{gaugetransformations}). Nevertheless,  by  introducing in
(\ref{gaugetransformations}) the following gauge parameters
\begin{equation}
\begin{split}
\varepsilon^{I}&=-\xi^{\mu}A_{\mu}^{I}, \\
\varepsilon_{i}^{I}&=\xi^{\mu}F_{\mu i}^{I},
\end{split}
\label{eq61}
\end{equation}
then, we obtain
\begin{equation}
A_{i}^{I}\rightarrow A_{i}^{I}+ \mathcal{L}_{\xi}A_{i}^{I}.
\end{equation}
Therefore, diffeomorphisms correspond to an internal symmetry of the
theory in the phase space. It is important to remark that  this
symmetry is devoid in ($\ref{5}$). In fact, in (\ref{5})  the gauge
transformations on the phase space correspond to $A\rightarrow A+
D\epsilon $ being different to the diffeomorphism symmetry
\cite{34}. We can also see that diffeomorphism symmetry implies
that the extended Hamiltonian (\ref{eq57}) is linear combination of
first class constraints
unlike  Yang-Mills theory where its Hamiltonian is not. In addition, all the information
obtained along this section has been performed with  the aim to know the local symmetries of the
theory under study, furthermore   will be useful
in future  works to study   the moduli space
of  the action (\ref{action BF}). It is important to observe, that the action (\ref{action BF}) and BF theory share the same
local symmetries, and  if the  coupling constant goes to zero, the action   (\ref{action BF}) and BF theory has  the same moduli space
as can be appreciated in (\ref{motioneq}). Nevertheless, if the coupling constant is not zero, (\ref{action BF}) and BF theory has the same local symmetries
but the moduli space will be different;  this issue will be studied in future works as well.\\
\newline
\newline
\newline
\noindent {\uno VII. Discussions}\vspace{.5em}\\
In this paper, the dimension of the moduli space for two and
four-dimensional BF theories valued in different gauge scenarios
have been determined using the Atiyah-Singer theorem . As an
important fact, we have used the connections that generate
simultaneously four-instantons and four-anti-instantons to
characterize the connections of a BF theory. This local information
allowed us to built the elliptic complex and then define its
corresponding moduli space. In addition we applied the results to
particular base manifolds and gauge bundles. On the other hand
within Dirac's method we have developed the Hamiltonian analysis of
a modified BF theory to obtain some significance results such as the
extended action, the extended Hamiltonian, the local degrees of
freedom and the gauge symmetries. As important results obtained
using the Hamiltonian method, we found that the theory is
diffeomorphisms covariant and with the use of the constraints we
concluded that it has zero physical local degrees of freedom, showing
in the appropriate limit only the global degrees of freedom
determined as the dimension of the moduli space. In this sense, in
the limit when the coupling constant goes to zero this modified
version of BF theory shares unlike the usual BF-YM theory, the same
moduli space and the same gauge transformations with the usual BF
theory. Nevertheless, the moduli space obtained for usual BF theory
and modified BF theory correspond to connections satisfying one of
the equations of motion, more precisely flat connections; but these
solutions are not invariant under all gauge transformations of the
theory as we have observed within Dirac's method. In order to take
into account this fact, we have to consider the second equation of
motion \ref{eq:BF} related to the field $B$, and build its
corresponding elliptic complex; this problem will have to be faced
within the setting of \textit{coupled} elliptic
complexes\cite{35}. In this manner the theory has been
characterized both globally and locally providing all necessary
elements to make progress in the quantization; these subjects will be reported in forthcoming works.

\begin{center}
{\uno ACKNOWLEDGMENTS}
\end{center}
This work was supported by the Sistema Nacional de Investigadores
and Conacyt (M\'{e}xico). The numerical analysis and graphics have been made using Mathematica.

\end{document}